\documentclass[preprint2]{aastex}

\shorttitle{Water ice and $^{12}$CO in V4332 Sgr}
\shortauthors{Banerjee, Varricatt and Ashok}
\begin{document}
\title{L $\&$ M band infrared studies of V4332 Sagittarii - detection of the
water-ice absorption band at 3.05 ${\rm{\mu}}$m and the CO fundamental band
in emission}

\author{Dipankar P.K. Banerjee}
\affil{Physical Research Laboratory, Navrangpura,  Ahmedabad 
Gujarat 380009, India}
\email{orion@prl.ernet.in}
\author{Watson P. Varricatt}
\affil{Joint Astronomy Center, 660 N. Aohoku Place, Hilo, Hawaii-96720, USA}
\email{w.varricatt@jach.hawaii.edu}
\and
\author{Nagarhalli M. Ashok}
\affil{Physical Research Laboratory, Navrangpura,  Ahmedabad 
Gujarat 380009, India}
\email{ashok@prl.ernet.in}

\begin{abstract}
L and M band observations of the nova-like variable V4332 Sgr are presented. 
Two significant results are obtained viz. the unusual detection of water ice
at 3.05${\rm{\mu}}$m  and  the fundamental band of $^{12}$CO at 4.67${\rm{\mu}}$m 
in emission.  The ice feature is a first detection in  a nova-like variable 
while the CO emission is rarely seen in novae. These results, when considered 
together with other existing data, imply that  V4332 Sgr could be a young 
object surrounded by a circumstellar disc containing gas, dust and ice. The 
reason for a nova-like outburst to occur in  such a system is unclear. But 
since planets are believed to form in such disks, it appears plausible that the 
enigmatic outburst of V4332 Sgr could be due to a planetary infall. We also 
give a more reliable estimate for an epoch of dust formation around V4332 Sgr
which appears to have taken place rather late in 1999 - nearly five years after 
its outburst.  
\end{abstract}


\section{Introduction}
We present here L and M band results on  V4332 Sgr. Recent studies of V4332 
Sgr have shown that it is an  interesting object and the present 
results further support this view. V4332 Sgr erupted in 1994 in a 
nova-like outburst with an outburst amplitude of 9.5 magnitude in the 
visible region. There was only one detailed study of the object 
during its outburst (Martini et al. 1999) which
showed that its eruption was different from that of a classical nova
or other classes of eruptive variables.  Interest in the object
has been rekindled  because of the recent outburst of V838 Mon which drew 
considerable attention due to its light-echo (Munari et al.
 2002; Bond et al. 2003). It is believed that V838 Mon, V4332 Sgr and M31 RV 
( a red-variable which exploded in M31 in 1988; Rich et al. 1989)
could be members of a  new class of eruptive objects (Munari et al. 2002; Bond
et al. 2003, Kimeswenger et al. 2002).  The cause
of the outburst in these objects does not appear to be satisfactorily
explained by conventional mechanisms. Thus new theories have been
proposed viz. a scenario involving the merger of main 
sequence stars (Soker $\&$ Tylenda, 2003) and a hypothesis invoking 
planetary-capture by an expanding star to explain the eruption 
(Retter $\&$ Marom, 2003). The present  data indicate that the 
second mechanism could be viable in  V4332 Sgr.

Recent infrared studies of V4332 Sgr have detected several bands of AlO 
at a low rotational temperature of 200-300K (Banerjee et al. 2003).
A considerable change in the spectral energy distribution (SED) of the 
object was  seen between 2MASS data of 1998 and observations in 2003 
indicating the formation of a dust shell between these two epochs 
(Banerjee et al. 2003). A better estimate of the epoch  when the
dust actually formed is discussed here. Optical spectroscopy of V4332 Sgr 
in 2003, showed an interesting  spectrum dominated by very 
strong emission in the resonance lines of KI and NaI (Banerjee $\&$ Ashok 
2004).  The SED of the star, derived from optical and IR data, indicated 
a central star with a black-body temperature of 3250K and an IR excess 
attributed to a  dust component at $\sim$ 900K (Banerjee $\&$ Ashok 2004). 

      \section{Observations}
      Observations  were done using the 3.8m UK
      Infrared Telescope (UKIRT). Spectroscopy was done using the UKIRT
      Imaging spectrometer (UIST), which uses different grisms to cover
      the 1.4-5 micron range. $L'$ (3.77 ${\rm{\mu}}$m) and $M'$ 
      (4.68 ${\rm{\mu}}$m) band photometry - not available earlier for 
      V4332 Sgr - was also done using UIST. Flat-fielding, spectral 
      calibration and other reduction procedures were done on the same lines 
      as our earlier $JHK$ study of V4332 Sgr (Banerjee et al. 2003).The log 
      of the observations and the observed $L', M'$ magnitudes of V4332 Sgr
      are given in Table 1. 
      
      \section{Results}
      \subsection{The water ice feature at 3.05 ${\rm{\mu}}$m and the 
      CO fundamental band emission} 
      Figure 1 shows the spectrum - the A-X bands of AlO in the $HK$ band, 
      reported earlier (Banerjee et al. 2003) are seen prominently in the present
      spectrum also  but are not discussed here.  A  remarkable feature - never 
      seen before in a nova-like object - is the deep, solid-state 3.05 
      ${\rm{\mu}}$m water-ice band formed due to the O-H stretching mode. 
      At very low temperatures, atoms and molecules can collide and adhere to 
      a dust grain  to
      produce an ice mantle on the surface. Atoms can migrate from one site to 
      another on the mantle to form a molecule - water ice is believed to form this
      way with H atoms combining with an O atom. The presence of cold water-ice 
      around V4332 Sgr is extremely unexpected since the ejecta of classical novae
      generally  evolve to high temperatures of $\sim$ 10${^{\rm 6}}$K (the coronal
      phase).  Following a standard procedure, we have obtained the optical depth 
      plot  of the ice feature by fitting a polynomial to the continuum around it 
      (Gibb et al. 2004).   The depth of the ice feature below this continuum
      was  found and   converted to an optical depth. The optical
      depth plot is shown in Figure 2. The 3.05 ${\rm{\mu}}$m feature was 
      compared with laboratory data for the optical depth of water-ice at different
      temperatures (10, 30, 50, 120 and 160K)  taken from the 
      Leiden database for ice analogs\footnote{ http://www.strw.leidenuniv.nl/$\sim$lab/}. 
      From a $\chi$${^{\rm 2}}$ test to the observed and model data, we find that the
      50K data gives the smallest value of $\chi$${^{\rm 2}}$. 
      The 30K data also gives a comparable value of $\chi$${^{\rm 2}}$  thus
      suggesting a low temperature   of $\sim$30-50K for the   water ice.  An 
      extended red wing between 3.3-3.8 ${\rm{\mu}}$m, which is not well fitted
      by the models, is  seen in the observed data. This extended ice wing is
      also  seen in several water ice detections but the species responsible for it 
      is unidentified (Gibb et al. 2004). From Figure 2,  the column 
      density of the  water-ice $N$ was calculated using $N$ = $\int$$\tau$d$\nu$/$A$, 
      where $A$ is the the band strength  for water-ice with a laboratory measured
      value of $A$ =  20$\times$10${^{\rm -17}}$  cm molecule${^{\rm -1}}$. While 
      carrying out the integration, we have  assumed that the missing data points 
      around 2.65 ${\rm{\mu}}$m (due to atmospheric cutoff), are  represented by 
      the data points of the 50K laboratory model in that region. We obtain a value 
      of $\int$$\tau$d$\nu$ =  362$\pm$27 cm${^{\rm -1}}$  leading to $N$ =
      (1.81$\pm$0.13)$\times$10${^{\rm 18}}$ cm${^{\rm -2}}$ - this value may be used
      in case of future detection of  other ices (CO${_{\rm 2}}$, 
      CH${_{\rm 3}}$OH, CH${_{\rm 4}}$ etc.) in V4332 Sgr to get a better 
      understanding of the   ice composition. 
       
      Another rare feature seen in V4332 Sgr is the  fundamental band ($\nu$ = 1-0)
      of $^{12}$CO at 4.67 ${\rm{\mu}}$m in emission.  There appear to be only a 
      few other detections of the CO fundamental  in emission - mostly towards 
      YSOs and Herbig AeBe stars (e.g. Blake $\&$ Boogert 2004; Pontoppidan et 
      al. 2002). In a few novae, emission in the CO first overtone bands has 
      been seen (Rudy et al. 2003 -  Table 3 therein; Evans et al. 1996) but the
      detection of the fundamental band appears rare (Lynch et al. 1997). The 
      expanded  CO ro-vibrational   spectrum  in V4332 Sgr is shown in Figure 3.
      Individual branch lines from P${_{\rm 1}}$ to P${_{\rm 12}}$ and
       R${_{\rm 0}}$ to R${_{\rm 12}}$
      are clearly seen. The lines are not resolved at their intrinsic width
      at the observed resolution of 1000.  A simple model for the $^{12}$CO
      emission was computed assuming LTE conditions with the level populations
      proportional to (2J+1)e${^{\rm -BhcJ(J+1)/kT}}$ where B and J are the 
      rotational constant for CO and  the rotational quantum number respectively; T 
      is the temperature. The strength of each P or R branch line was then
      obtained using the transition probabilities for the lines given by 
      Goorvitch (1994).  The line positions were also taken from Goorvitch (1994).
      The model spectrum thus obtained was  convolved with
      a Gaussian instrument function of FWHM  0.046 ${\rm{\mu}}$m (i.e. a resolution
      of 1000). From the model fits, it is difficult to constrain the temperature 
      of the gas too accurately, but a relatively low temperature of $\sim$300-400K
      is suggested. This could possibly be the reason  for the  absence in V4332 Sgr
      of the   first overtone CO bands at 2.3-2.45${\rm{\mu}}$m  which need a 
      higher temperature for excitation. These bands have been modeled to have
      a   temperature in the range 2500-4500K in novae (Rudy et al. 2003 ; Evans
      et al. 1996).  Further, the strong presence of AlO bands in V4332 Sgr in 
      the same  spectral region as the  CO first overtone bands, makes it difficult 
      to draw definite conclusions on  the absence/presence of the latter. A better
      estimate for the CO temperature  than that  derived here would require  
      modeling based on higher resolution  spectra (e.g. Blake $\&$ Boogert 2004; 
      Pontoppidan et al. 2002).

      \subsection{The case for V4332 Sgr being a young object with a surrounding
       circumstellar disk}
       The detection of water ice at 30-50K in V4332 Sgr is very intriguing 
      since such a low temperature component  is not expected in novae 
      ejecta.  Based on the recent
      comprehensive survey (Gibb et al. 2004 and  references therein), most
      water-ice detections are seen towards embedded YSOs/protostars. 
      Such objects are known to have circumstellar disks (CSDs) around them. 
      In addition, water-ice has also been seen in Herbig AeBe stars
      and a few  T Tauri stars (Creech Eakman et al. 2002; Meeus et al. 2001,
      Gibb et al. 2004). These are all young  objects  observationally known
      to possess CSDs.  Furthermore, all the  gaseous components of V4332 
      Sgr are seen in emission i.e. the   KI/NaI, CO, AlO and TiO lines (Banerjee 
      $\&$ Ashok 2004; Banerjee et al. 2003). These species are at a low 
      temperature ( 200K for AlO ; a low excitation temperature is also implied
      for the KI/NaI lines which need only $\sim$ 1.5-2eV for excitation). If
      the region in which these species exist were to be a shell surrounding
      the central 3250K star of V4332 Sgr (Banerjee $\&$ Ashok 2004), the 
      relatively cold gas of these species should lead to lines in absorption.
      Such is not the case. Instead, if the various species are in a disk, their
      lines can be  expected in emission. Furthermore, in the case of a disk,
      the central star would also be observable since it will be unobscured
      (some obscuration could occur depending on the orientation and
      thickness of the disk). Such is the case in V4332 Sgr where the
      continuum from the central star is clearly seen.   Additional 
      support for a CSD is the infrared excess seen in V4332 
      Sgr (Banerjee $\&$ Ashok 2004). This IR excess can be attributed to dust in
      the disk and not in a shell around the star since the dust shell
      would obscure the visible radiation of the star. A similar reasoning
      is used to explain the IR excess in T Tauri stars. It is
      also relevant to note that many of the known CO fundamental detections are in
      disk-dominated Herbig Ae stars (Blake $\&$ Boogert 2004). The considerable 
       width of the KI       lines in V4332 Sgr was also
      shown to be consistent with  line broadening arising
      from rotational motion of gas in a disk (Banerjee $\&$ Ashok 2004). 
      An additional signature      that V4332 Sgr has pre-existing matter
      around it at the time of the 1994 outburst  comes from
      the high-resolution H$\alpha$ profiles of the object taken shortly 
      after its outburst.  These profiles (Figure 8 of Martini et al. 1999) 
      show a deep absorption  trough at the center which could be caused by
      absorption by  pre-existing matter. Therefore,  many of the observed
      characteristics of V4332 Sgr support the existence of a   CSD around the  
      object. However, it is difficult  to rule out the possibility of  the
      matter being in clumps instead of a disk. It may be pointed out that the
      maximum value of $\tau$ $\sim$ ~ 1 of the ice feature (Figure 2), would 
      imply a large extinction of A${_{\rm v}}$ $\sim$ 14 magnitudes towards 
      the central star, had the 3.05${\rm{\mu}}$m absorption arisen because of
      cold, intervening matter in the line of sight ( as  in  embedded YSOs in
       dark clouds; Whittet 1992). As the central star  appears unobscured
      (V= 17.52, Banerjee $\&$ Ashok 2004), the 3.05${\rm{\mu}}$m feature could 
      instead be caused by ice in the CSD  viewed against the IR flux from dust in
      the disc  leading  to the absorption feature.
           
      \subsection{The formation of dust around V4332 Sgr}
      It was  earlier shown that the $JHK$ fluxes of V4332 Sgr changed 
      considerably between 2MASS observations of May, 1998 ($J$ =12.1, $H$ = 
      11.6, $K$ = 10.992) and observations of June, 2003 ($J$ =13.25, $H$ = 
      11.986, $K$ = 10.023; Banerjee et al. 2003).
      While the 2 MASS data showed no IR excess and was well modeled
      by a 3250K blackbody, the 2003 data showed a composite SED of a
      blackbody component at 900K (due to newly formed dust) and
      a  weakening of the 3250K component due to obscuration by
      dust. However, it was not possible to establish with accuracy at which 
      stage between 1998 and 2003 the dust formed. Data from the DENIS survey
      is now available and this shows that in September, 1999 V4332 Sgr had 
      IR magnitudes of $J$= 12.46 and $K$ = 10.65 (DENIS did not observe in $H$ band).
      A comparison between 2MASS and DENIS $JK$ magnitudes clearly shows
      the IR flux increasing towards  longer wavelengths indicating that
      dust  formation had begun by the DENIS epoch. The formation of dust
      therefore took place at t${_{\rm dust}}$ $\sim$ 5.5 years after the outburst
      of February 1994. This is much larger than the typical   timescale of 50-100 
      days for dust formation in classical novae. The difficulty in having a 
      large dust-condensation time in novae is that the density in the  ejecta 
      decreases to a low level - due to expansion -   that does not favor 
      dust-formation. The density $\rho$ in the ejecta of novae that form dust 
      is in the range 10${^{\rm -15}}$ to 10${^{\rm -16}}$gcm${^{\rm -3}}$ (Gehrz 1988). 
      In V4332 Sgr, the electron density  n${_{\rm e}}$ after $\sim$ 2-3 months 
      from the outburst was  estimated to be 10${^{\rm 8-9}}$ cm${^{\rm -3}}$ 
      (Martini et al. 1999). 
      A geometric dilution for the expanding ejecta matter  may be a reasonable 
      assumption i.e n${_{\rm e}}$ $\alpha$ 1/r${^{\rm 2}}$ provided 
      the ejecta expands freely, unhindered by pre-existing
      matter. Assuming n${_{\rm e}}$ $\alpha$ 1/r${^{\rm 2}}$, 
      after a     time of 5.5yr, n${_{\rm e}}$ should have evolved to $\sim$ 
      10${^{\rm 5-6}}$ cm${^{\rm -3}}$  giving a  value of 
      $\rho$ = 10${^{\rm -17}}$ - 10${^{\rm -18}}$gcm${^{\rm -3}}$. Such a low 
      value of $\rho$ appears unfavorable for  dust formation.  It therefore
      appears that dust formation may not have occured per se in the ejecta of the 1994 
      outburst. It is  likely that the ejecta, after a stage of free expansion,
      could have impinged on pre-existing matter around V4332 Sgr - sufficiently 
      far from the star -  thereby heating it  and giving rise to the observed 
      emission at present from different species.  This could be one possible 
      scenario to explain the large observed time for the formation of dust. In
      a connected aspect, we note that  the present $L'$, $M'$ data of Figure 1, shows 
      the continuum peaking at     $\sim$ 4.4${\rm{\mu}}$m indicating 
      a lower, dust temperature than the 900K that was inferred earlier. 
      However, mid/far-IR observations planned on  the Spitzer Telescope 
      should give  more definite information on the dust temperature. 
      
     \section{Discussion: Is a planetary infall responsible for the outburst in 
      V4332 Sgr?}       
      More than a hundred stars with planets (SWPs) are known 
      today\footnote{http://cfa-www.harvard.edu/planets/cat1.html} and in
      $\sim$15$\%$ of these the planetary companion is unexpectedly
      close to the host star ( $\le$ 0.05 AU).  The orbits of such
      short-period planets around sun-like stars can be unstable because of
      tidal dissipation and they can be subsequently consumed by the host
      stars (Sandquist et al. 2002; Rasio et al. 1996).  More evolved stars are 
      also capable of swallowing their planets as they expand (Siess $\&$ Livio 
      1999).  The viability of planetary ingestion is observationally supported 
      by the enhanced metallicity seen in many SWPs - and the presence of Li 
      isotopes in one of them - suggesting that planetary 
      material has been  accreted by the host star in the past (Santos et al. 2000; 
      Gonzalez et al. 2001; Israelian et al. 2003). In particular context of V4332 
      Sgr type of objects, Retter and Marom (2003) have explained the multi-peak 
      light curve of V838 Mon as due to  an expanding star ingesting its planets.
      
      We have shown that there is considerable evidence that V4332 Sgr is 
      surrounded by a    cold, dusty disk. It is precisely in such an 
      environment that planets are believed to be born  from the coagulation
       and accretion of solid ice and dust particles (Lissauer 1993).
      Therefore, while it may appear speculative, it is  not entirely unlikely
      that a planet existed around V4332 Sgr and its infall led to the outburst.
       The   plausibility of this scenario is strengthened since other
      conventional mechanisms - such as a thermo-nuclear runaway on a white
      dwarf surface (invoked for classical, recurrent, symbiotic novae) or
      a final Helium shell flash (invoked for a born-again AGB star) - fail
      to explain the pre- and post-outburst properties of V4332
      Sgr or similar objects. It was shown (Retter $\&$ Marom 2003)  in the 
      analysis for V838 Mon, that the  gravitational energy released by a 
      1M${_{\rm J}}$ planet falling into a solar-mass star is 
      4$\times$10${^{\rm 5}}$L$_\odot$.  The outburst luminosity of V4332 Sgr
      is difficult to estimate as the distance $d$ to the object is very
      uncertain. Using an estimate of $d$ = 300 pc (Martini et al. 1999) we 
      have shown that the   star has a quiescent luminosity of ~ 0.3$L$$_\odot$
      (Banerjee et al. 2003). A 9.5 magnitude brightening at outburst would 
      yield an outburst luminosity of $\sim$  2$\times$10${^{\rm 3}}$L$_\odot$. 
      This is lower than the predicted energy       release for the
      capture of a 1M${_{\rm J}}$ planet but could be made consistent by revising 
      the distance (the current distance estimate to V838 Mon has been  revised
      upwards by a factor of nearly 10 from its first estimate), invoking a 
      smaller mass for the captured planet or  that a large part of the released 
      gravitational energy has gone into  expansion of the stellar envelope 
      (Retter $\&$ Marom 2003) rather than visible radiation. However, we feel 
      there are many aspects about the nature of V4332 Sgr that are still unclear 
      and there is scope for alternative models, apart from that proposed here, to
      explain its outburst.  
       
 \begin{acknowledgements}
Research at the PRL is funded by the Dept. of Space, Govt. of India.  We thank 
the UKIRT service program for observation time  covering  a part of these
observations. UKIRT is operated by the JAC, Hawaii on behalf of the UK 
PPARC. We are grateful to the referee for several helpful comments. 
	  
\end{acknowledgements}

%______________________________________________________________

\clearpage
\begin{figure}
\plotone{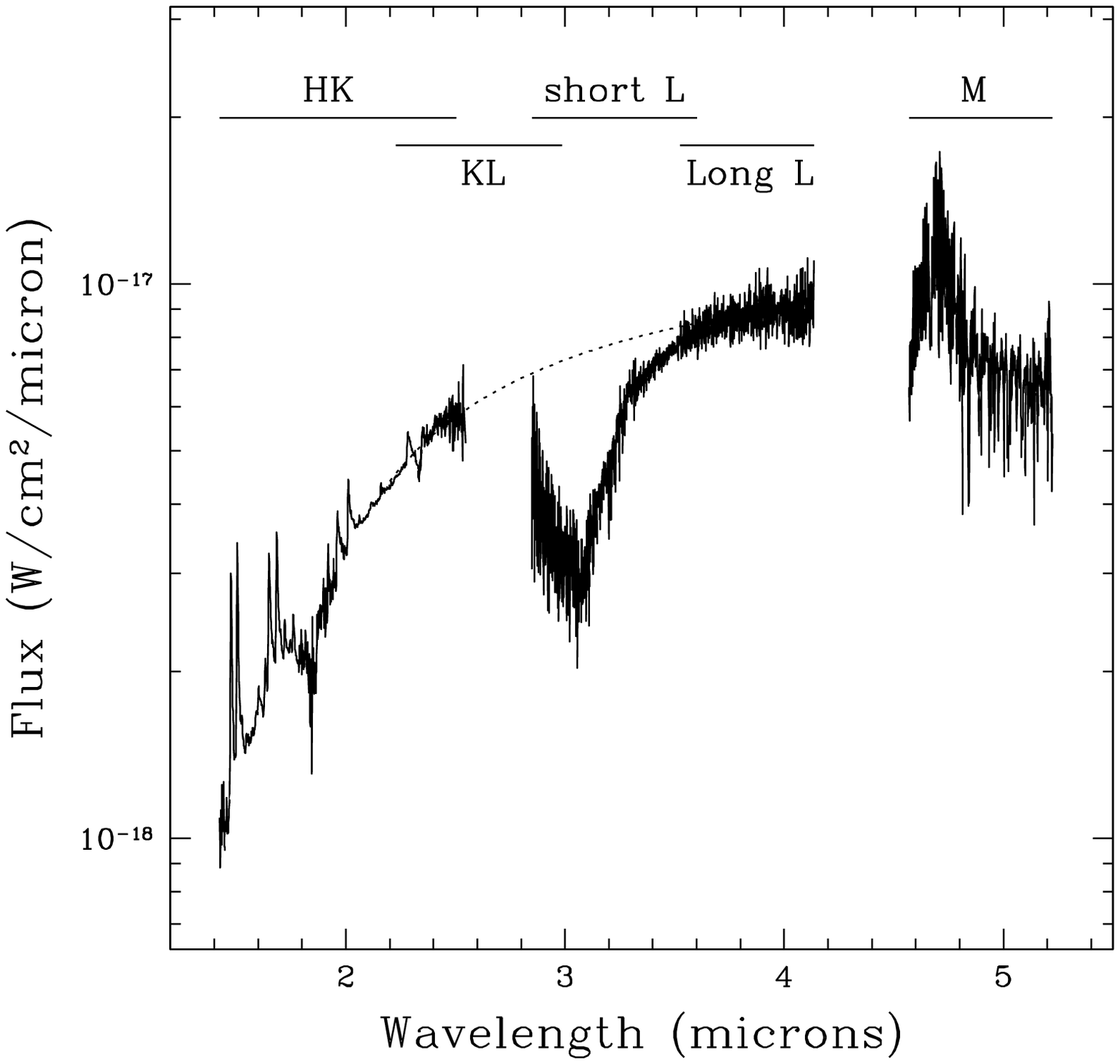}
\caption{ The near-IR spectrum of V4332 Sgr  is shown by bold line. The  
extent of the individual spectra - used  to obtain  the composite spectrum - 
are marked with straight lines on top. The spectra have been flux calibrated 
using $L', M'$ magnitudes from the present work  and  previously measured 
$H, K$  magnitudes.  A continuum fit between 2.2 and 4.13 ${\rm{\mu}}$m to
 obtain the optical depth of the 3.05 ${\rm{\mu}}$m ice feature is  shown 
by a dotted line. Some telluric water vapor lines between 4.8 to 5.2 
${\rm{\mu}}$m (this region has poor atmospheric transmission) were seen in
the spectra which have been removed. These were not felt to be intrinsic to 
the source but due to imperfect    cancellation  while dividing the object 
and  standard star spectra. The gaps in the spectra at  2.65 and 4.4 
${\rm{\mu}}$m are due to poor atmospheric transmission in these regions. 
\label{fig1}}
\end{figure}

\clearpage
\begin{figure}
\plotone{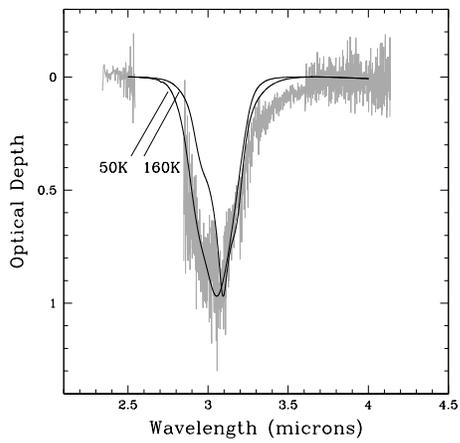}
\caption{ The observed optical depth plot of the 3.05 ${\rm{\mu}}$m ice 
feature is shown by  grey line. The black lines show laboratory models for 
water-ice at 50 and 160K. The shape of the observed 3.05 ${\rm{\mu}}$m feature
is better represented  by the 50K curve (and also with a 30K model fit 
that was tried) rather than higher temperature fits (e.g. the 160K curve). 
\label{fig2}}
\end{figure}

\clearpage
\begin{figure}
\plotone{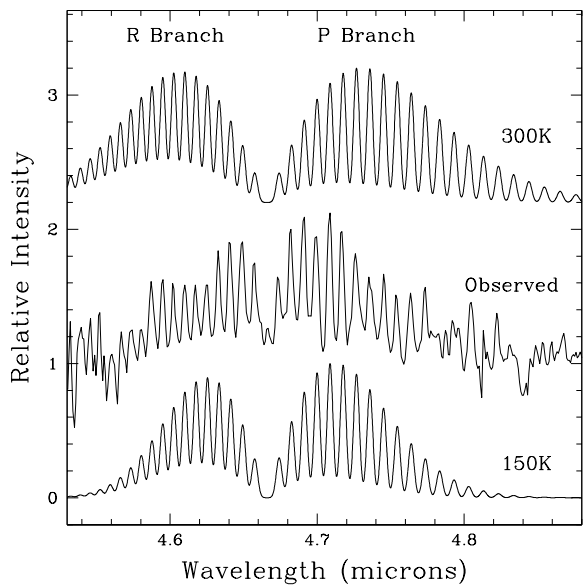}
\caption{Model and observed plots for the $^{12}$CO fundamental emission in 
V4332 Sgr.  Though the S/N of the spectrum is slightly inadequate,  the 
relative strengths of the  rotational lines of the R and P branches are 
better reproduced by the 300K model than the 150K model.  This is 
specially so for the higher R and P branch lines which become  weak for
the 150K model fit. \label{fig3}}
\end{figure}

\clearpage
\begin{table}
\caption{Log of  observations for V4332 Sgr. }
\begin{tabular}{cccccc}
\hline 
\hline\\
 Date(UT)            & Band    &  Resol- & Integration&  Standard & Magni-\\
                     &         &  ution  & time(s)    &  Star     & tude  \\
&&&&&\\ 
SPECTROSCOPY&&&&&\\
  2003 Sept. 5.294   & M       &  1000   &      312   &  BS 6998  &       \\
  2003 Sept. 5.359   & KL      &  700    &      360   &  BS 6998  &       \\
  2003 Sept. 5.328   & Short L &  650    &      720   &  BS 6969  &       \\
  2004 April 15.606  & HK      &  500	 &      720   &  BS 7038  &       \\
  2004 April 29.642  & Long L  &  1150	 &      320   &  BS 6998  &       \\
&&&&&\\
PHOTOMETRY&&&&&\\
2003 Sept. 5.262     & L'      &         &      252   &  HD 161743& 7.23 $\pm$ 0.025\\
2003 Sept. 5.276     & M'      &         &      322   &  HD 161743& 6.10 $\pm$ 0.03\\
\hline
\hline\\
\end{tabular} 
\end{table}
\end{document}